\documentclass[a4paper,10pt]{article}
\usepackage{amsfonts, amsthm, amsmath}

\numberwithin{equation}{section}
\title{ {\bf
Integrable and superintegrable systems with spin in three-dimensional 
Euclidean space }}

\author{\vspace{1cm}\\
         {\bf Pavel Winternitz}$^{1,2}$
         \thanks{E-mail address:
        wintern@crm.umontreal.ca}
         {\,\,and \bf \.{I}smet Yurdu\c{s}en}$^1$
        \thanks{E-mail address:
       yurdusen@crm.umontreal.ca}
\\
\\$^1$Centre de Recherches Math\'{e}matiques, 
Universit\'{e} de Montr\'{e}al,\\ CP 6128, Succ. Centre-Ville, Montr\'{e}al, 
Quebec H3C 3J7, Canada
\\
\\$^2$D\'{e}partement de Math\'{e}matiques et de Statistique, 
Universit\'{e} de Montr\'{e}al,\\ CP 6128, Succ. Centre-Ville, Montr\'{e}al, 
Quebec H3C 3J7, Canada}

\date{\today}
\newpage
\begin{document}
\setlength{\baselineskip}{24pt} 
\maketitle
\setlength{\baselineskip}{7mm}
\begin{abstract}
A systematic search for superintegrable quantum Hamiltonians describing 
the interaction between two particles with spin $0$ and $\frac{1}{2}$, 
is performed. We restrict to integrals of motion that are first-order 
(matrix) polynomials in the components of linear momentum. Several such 
systems are found and for one non-trivial example we show how 
superintegrability leads to exact solvability: we obtain exact 
(nonperturbative) bound state energy formulas and exact 
expressions for the wave functions in terms of products of 
Laguerre and Jacobi polynomials.
\end{abstract}


\section{Introduction}\label{intro}
The purpose of this research program is to perform 
a systematic study of integrability and superintegrability 
in the interaction of two particles with spin. Specifically 
in this article we consider a system of two nonrelativistic 
particles, one with spin $s=\frac{1}{2}$ (e.g. a nucleon) 
the other with spin $s=0$ (e.g. a pion), moving in the 
three-dimensional Euclidean space $E_3$.

The Pauli-Schr\"{o}dinger equation in this case will have the form 
\begin{eqnarray}
H\Psi=\left(-\frac{\hbar^2}{2} \Delta + V_0(\vec{x}) + \frac{1}{2}\Big\{V_1(\vec{x}), 
(\vec{\sigma}, \vec{L})\Big\}\right)\Psi = E\Psi\,,
\label{generalhamiltonianwithhsec1}
\end{eqnarray}
where the $V_1(\vec{x})$ term represents the spin-orbital interaction. 
We use the notation 
\begin{eqnarray}
p_1=-i \hbar \partial_x, \qquad p_2=-i \hbar \partial_y, \qquad p_3=-i \hbar \partial_z\,,
\label{defmomentumlar} 
\end{eqnarray}
\begin{eqnarray}
L_1=i \hbar (z\partial_y-y\partial_z),\qquad L_2=i \hbar (x\partial_z-z\partial_x), 
\qquad L_3=i \hbar (y\partial_x-x\partial_y)\,,
\label{defangularmomentumlar} 
\end{eqnarray}
\begin{eqnarray}
\sigma_1=\left(\begin{array}{cc} 0& 1\\ 1& 0 \end{array}\right), 
\qquad 
\sigma_2=\left(\begin{array}{cc} 0& -i\\ i& 0 \end{array}\right), 
\qquad 
\sigma_3=\left(\begin{array}{cc} 1& 0\\ 0& -1 \end{array}\right)\,,
\label{defsigmamatrices} 
\end{eqnarray}
for the linear momentum, angular momentum and Pauli matrices, 
respectively. The curly bracket in 
(\ref{generalhamiltonianwithhsec1}) denotes an anticommutator. 

For spinless particles the Hamiltonian 
\begin{eqnarray}
H=-\frac{\hbar^2}{2} \Delta + V_0(\vec{x})\,,
\label{spinlesshamiltonian}
\end{eqnarray}
is a scalar operator, whereas $H$ in (\ref{generalhamiltonianwithhsec1}) 
is a $2 \times 2$ matrix operator and $\Psi(\vec{x})$ is a two 
component spinor. 

In the spinless case (\ref{spinlesshamiltonian}) the Hamiltonian 
is integrable if there exists a pair of commuting integrals of 
motion $X_1$, $X_2$ that are well-defined quantum mechanical 
operators, such that $H$, $X_1$ and $X_2$ are algebraically independent. 
If further algebraically independent integrals $Y_i$ exist, the system is 
superintegrable. The best known superintegrable systems in $E_3$ are 
the hydrogen atom and the harmonic oscillator. Each of them is maximally 
superintegrable with $2n-1=5$ independent integrals, generating an 
$o(4)$ and an $su(3)$ algebra, respectively \cite{Fock, Jauch}. 

A systematic search for quantum and classical superintegrable 
scalar potentials in (\ref{spinlesshamiltonian}) with 
integrals that are first- and second-order polynomials in 
the momenta was performed some time ago 
\cite{Makarov, Evans.a, Evans.b}. First-order integrals 
correspond to geometrical symmetries of the potential, 
second-order ones are directly related to the separation 
of variables in the Schr\"{o}dinger equation 
or Hamilton-Jacobi equation in the classical case 
\cite{Makarov, Fris, Miller, Kalnins}. 

First- and second-order integrals of motion are 
rather easy to find for Hamiltonians of the type 
(\ref{spinlesshamiltonian}) in Euclidean space. 
The situation with third- and higher-order integrals 
is much more difficult \cite{Drach, Gravel.a, Gravel.b}. 

If a vector potential term, corresponding e.g. to a 
magnetic field is added, the problem becomes much more 
difficult and the existence of second-order integrals 
no longer implies the separation of variables 
\cite{Charest, Pucacco}. 

The case (\ref{generalhamiltonianwithhsec1}) with a 
spin-orbital interaction turns out to be quite rich 
and rather difficult to treat systematically. 
In a previous article we have considered the same 
problem in $E_2$ \cite{Winternitz2}. Here we 
concentrate on the Hamiltonian 
(\ref{generalhamiltonianwithhsec1}) in $E_3$ but 
restrict to first-order integrals. Thus we search for 
integrals of motion of the form 
\begin{eqnarray}
X=(A_0+\vec{A} \cdot \vec{\sigma})p_1+(B_0+\vec{B} \cdot \vec{\sigma})p_2+
(C_0+\vec{C} \cdot \vec{\sigma})p_3+\phi_0+\vec{\phi}\,\cdot \vec{\sigma} 
\nonumber \\
-\frac{i \hbar}{2}\left\{(A_0+\vec{A} \cdot \vec{\sigma})_x
+(B_0+\vec{B} \cdot \vec{\sigma})_y+(C_0+\vec{C} \cdot \vec{\sigma})_z
\right\}
\,, \label{appendixmostgeneralintofmotion}
\end{eqnarray}
where $A_0$, $B_0$, $C_0$, $\phi_0$ and $A_i$, $B_i$, $C_i$, $\phi_i$ 
($i=1,2,3$) are all scalar functions of $\vec{x}$.

In Section \ref{gauge} we show that a spin-orbital 
interaction of the form
\begin{equation}
V_1 = \frac{\hbar}{r^2}\,,
\label{gaugeinducedpotintr}
\end{equation}
can be induced by a gauge transformation from a 
purely scalar potential $V_0(\vec{x})$ (in particular 
from $V_0 = 0$). In Section \ref{determineq} we derive 
and discuss the determining equations for the 
existence of first-order integrals. In Section \ref{o3multiplets} 
we restrict to rotationally invariant potentials $V_0(r)$ and 
$V_1(r)$ and classify the integrals of motion into 
$O(3)$ multiplets. Solutions of the determining equations are 
obtained in Section \ref{solutionsofcommeq}. Superintegrable 
potentials are discussed in Section \ref{results}. 
In Section \ref{example} we solve 
the Pauli-Schr\"{o}dinger equation for one superintegrable system 
explicitly and exactly. Finally, the conclusions 
and outlook are given in Section \ref{conclusion}.


\section{Spin-orbital interaction induced by a gauge transformation}
\label{gauge}
\subsection{Derivation} 
In this Subsection we show that a spin-orbit term could be gauge 
induced from a scalar Hamiltonian (\ref{spinlesshamiltonian}) 
by a gauge transformation. The transformation matrix must be an 
element of $U(2)$
\begin{equation}
U=e^{i\beta_4}\left(\begin{array}{cc} e^{i\beta_1}\cos(\beta_3)& e^{i\beta_2}\sin(\beta_3)\\ -e^{-i\beta_2}\sin(\beta_3)& e^{-i\beta_1}\cos(\beta_3) \end{array}\right)\,, \label{gaugetrfmatrix}
\end{equation}
where $\beta_j$ ($j=1,2,3,4$) are real functions of ($x, y, z$). It 
is seen that in order to generate a spin-orbit term we need to have 
\begin{equation}
\hbar^2 \, U^{\dagger} (\vec{\nabla} U)\cdot \vec{\nabla} = \Gamma\,\, (\vec{\sigma}, \vec{L})\,,
\label{generatsigmaL}
\end{equation}
where $\Gamma$ is an arbitrary real scalar function of ($x, y, z$). Equation 
(\ref{generatsigmaL}) implies $12$ first-order partial differential 
equations for $\beta_j$ and $\Gamma$, three of which are $\beta_{4,k} = 0$, 
$k=1,2,3$. Hence, without loss of generality we choose $\beta_{4} = 0$ 
and then write the remaining $9$ equations as
\begin{eqnarray}
&&\cos^2(\beta_3)\beta_{1,z} - \sin^2(\beta_3)\beta_{2,z} = 0\,, \label{gaugedeteq1} \\
&&\cos^2(\beta_3)\beta_{1,x} - \sin^2(\beta_3)\beta_{2,x} = 0\,, \label{gaugedeteq2} \\
&&\hbar \Big(\cos^2(\beta_3)\beta_{1,y} - \sin^2(\beta_3)\beta_{2,y}\Big) = -x \Gamma\,, \label{gaugedeteq3} \\
&&\hbar \Big(\cos(\beta_2 - \beta_1) \beta_{3,x} -\frac{1}{2} \sin(\beta_2 - \beta_1) \sin(2\beta_3) (\beta_2 + \beta_1)_x\Big) = -z \Gamma\,, \label{gaugedeteq4} \\
&&\hbar \Big(\cos(\beta_2 - \beta_1) \beta_{3,z} -\frac{1}{2} \sin(\beta_2 - \beta_1) \sin(2\beta_3) (\beta_2 + \beta_1)_z\Big) = x \Gamma\,,\label{gaugedeteq5}  \\
&&\hbar \Big(\sin(\beta_2 - \beta_1) \beta_{3,y} +\frac{1}{2} \cos(\beta_2 - \beta_1) \sin(2\beta_3) (\beta_2 + \beta_1)_y\Big) = z \Gamma\,, \label{gaugedeteq6} \\
&&\hbar \Big(\sin(\beta_2 - \beta_1) \beta_{3,z} +\frac{1}{2} \cos(\beta_2 - \beta_1) \sin(2\beta_3) (\beta_2 + \beta_1)_z\Big) = -y \Gamma\,, \label{gaugedeteq7} \\
&&\cos(\beta_2 - \beta_1) \beta_{3,y} -\frac{1}{2} \sin(\beta_2 - \beta_1) \sin(2\beta_3) (\beta_2 + \beta_1)_y = 0\,, \label{gaugedeteq8} \\
&&\sin(\beta_2 - \beta_1) \beta_{3,x} +\frac{1}{2} \cos(\beta_2 - \beta_1) \sin(2\beta_3) (\beta_2 + \beta_1)_x = 0\,.\label{gaugedeteq9} 
\end{eqnarray}

From equations (\ref{gaugedeteq4})-(\ref{gaugedeteq9}) we obtain 
\begin{eqnarray}
&&\beta_j = \beta_j(\xi,\eta)\,, \qquad j=1,2,3\, \qquad {\rm where} \qquad \xi=\frac{x}{z}\,,  \quad \eta=\frac{y}{z}\,, \nonumber \\
&&\cos(\beta_2 - \beta_1) = -\frac{\beta_{3,\xi}}{\sqrt{\beta_{3,\xi}^{\,2} + \beta_{3,\eta}^{\,2}}}\,, \qquad 
\sin(\beta_2 - \beta_1) = \frac{\beta_{3,\eta}}{\sqrt{\beta_{3,\xi}^{\,2} + \beta_{3,\eta}^{\,2}}}\,, \nonumber \\
&&(\beta_2 + \beta_1)_{\xi} = \frac{2\beta_{3,\eta}}{\sin(2\beta_3)}\,, \qquad 
(\beta_2 + \beta_1)_{\eta} = -\frac{2\beta_{3,\xi}}{\sin(2\beta_3)}\,, 
\label{secondsetgaugedeteq1}
\end{eqnarray}
and 
\begin{eqnarray}
\Gamma = \hbar \frac{\sqrt{\beta_{3,\xi}^{\,2} + \beta_{3,\eta}^{\,2}}}{z^2}\,.
\label{secondsetgaugedeteq2}
\end{eqnarray}

Introducing $\beta_j(\xi,\eta)$ into the equations 
(\ref{gaugedeteq1})-(\ref{gaugedeteq3}) and using the 
compatibility condition for the mixed partial derivatives 
of $(\beta_2 + \beta_1)$ we obtain the following 
three partial differential 
equations 
\begin{eqnarray}
&\,&\beta_{3, \xi \xi} + \beta_{3, \eta \eta} = 2(\beta_{3,\xi}^{\,2} + \beta_{3,\eta}^{\,2}) \cot(2\beta_3)\,, \nonumber \\
&\,&\beta_{3, \eta} \beta_{3, \xi \eta} - \beta_{3, \xi} \beta_{3, \eta \eta} =  2(\beta_{3,\xi}^{\,2} + \beta_{3,\eta}^{\,2})\left(\xi\sqrt{\beta_{3,\xi}^{\,2} + \beta_{3,\eta}^{\,2}}-\cot(2\beta_3)\beta_{3, \xi}
\right)\,, \nonumber \\
&-&\!\!\!\beta_{3, \eta} \beta_{3, \xi \xi} + \beta_{3, \xi} \beta_{3, \xi \eta} =  2(\beta_{3,\xi}^{\,2} + \beta_{3,\eta}^{\,2})\left(\eta\sqrt{\beta_{3,\xi}^{\,2} + \beta_{3,\eta}^{\,2}}-\cot(2\beta_3)\beta_{3, \eta}
\right)\,,
\end{eqnarray}
which could be solved for the highest-order derivatives of $\beta_3$ 
(i.e. $\beta_{3, \xi \xi}$, $\beta_{3, \xi \eta}$ and 
$\beta_{3, \eta \eta}$). Then, the compatibility conditions of these give 
\begin{equation}
\sqrt{\beta_{3,\xi}^{\,2} + \beta_{3,\eta}^{\,2}} = -\frac{1}{\xi^2+\eta^2+1}\,,
\end{equation}
which implies that $\Gamma=-\frac{\hbar}{r^2}$. Hence, we conclude that 
$V_1 = \frac{\hbar}{r^2}$ is gauge induced and it is the only potential 
which could be generated from a scalar Hamiltonian by a gauge transformation.  

The explicit form of the gauge transformation $U$ is found as
\begin{equation}
\beta_1=\varphi + c_1\,, \quad \beta_2=c_2\,, \quad \beta_3=-\theta + c_3\,, \quad \beta_4=0\,, \quad 0\le\theta\le \pi\,, \quad 0\le\varphi<2\pi\,,
\label{specifiedgaugetrfmatrix}
\end{equation}
where $c_1$, $c_2$ and $c_3$ are the following constants
\begin{equation}
c_1 = c_2 \pm \pi\,, \qquad {\rm and} \qquad c_3 = \pm \frac{\pi}{2}\, 
\end{equation}
and $\theta$, $\varphi$ are the spherical coordinates.

With this transformation matrix the transformed Hamiltonian is found to be
\begin{equation}
\tilde H=U^{-1}\Bigg(-\frac{\hbar^2}{2} \Delta + V_0(\vec{x})\Bigg)U=-\frac{\hbar^2}{2} \Delta + V_0(\vec{x}) +\frac{\hbar^2}{r^2}+\frac{\hbar}{r^2}\vec{\sigma}\cdot\vec{L}\,. \label{gaugetransformedhamiltonian}
\end{equation}

\subsection{Integrals for $V_0=V_0(r)$ and $V_1=\frac{\hbar}{r^2}$}\label{subgt1}
The potential $V_1=\frac{\hbar}{r^2}$ is gauge induced from a 
Hamiltonian of the form (\ref{spinlesshamiltonian}) (though 
each term is multiplied by a $2 \times 2$ identity matrix). 
Hence the integrals for this 
case are just the gauge transforms of the integrals of motion of this 
Hamiltonian (i.e. $L_j$ and $\sigma_j$). They can be written as 
\begin{eqnarray}
J_i=L_i+\frac{\hbar}{2}\sigma_i\,,  &\,& \qquad
S_i=-\frac{\hbar}{2}\sigma_i+\hbar \frac{x_i}{r^2}(\vec{r},\vec{\sigma})
\,, \label{generatorsforsixdimalg}
\end{eqnarray}
and satisfy the following commutation relations
\begin{eqnarray}
[J_i, J_j]=i \hbar \epsilon_{ijk}J_k\,, 
\quad [S_i, S_j]=i \hbar \epsilon_{ijk}S_k\,, 
\quad [J_i, S_j]=i \hbar \epsilon_{ijk}S_k\,.
\label{commutationrelationso3}
\end{eqnarray}
The Lie algebra $\mathcal{L}$ is isomorphic to a direct sum 
of the algebra $o(3)$ with itself
\begin{eqnarray}
\mathcal{L} \sim {o}(3)\oplus {o}(3) = \{\vec{J} - \vec{S}\}
\oplus \{\vec{S}\}
\,.\label{algdirectsumo3o3}
\end{eqnarray}

\subsection{Integrals for $V_0=\frac{\hbar^2}{r^2}$ and $V_1=\frac{\hbar}{r^2}$}\label{subgt2}
Since these potentials are gauge induced from a free Hamiltonian, 
the integrals are just the gauge transforms of $L_j$, $p_j$ and 
$\sigma_j$, which can be written as
\begin{eqnarray}
J_i=L_i+\frac{\hbar}{2}\sigma_i\,,  &\,& \qquad
\Pi_i=p_i-\frac{\hbar}{r^2}\epsilon_{ikl}x_k\sigma_l\,, \nonumber \\
S_i&=&-\frac{\hbar}{2}\sigma_i+ \hbar \frac{x_i}{r^2}(\vec{r},\vec{\sigma})
\,. \label{generatorsforninedimalg}
\end{eqnarray}
They satisfy the following commutation relations
\begin{eqnarray}
[J_i-S_i, S_j]=0\,, \qquad [\Pi_i, S_j]=0\,, \qquad [\Pi_i, \Pi_j]=0\,, 
\nonumber \\
\,[J_i-S_i, J_j-S_j]=i \hbar \epsilon_{ijk}(J_k-S_k)\,, 
\quad [J_i-S_i, \Pi_j]=i \hbar \epsilon_{ijk}\Pi_k
\,. \label{commutationrelationsforoneoverrsqr}
\end{eqnarray}
Hence the $9$-dimensional Lie algebra $\mathcal{L}$ is isomorphic to a direct sum 
of the Euclidean Lie algebra $e(3)$ with the algebra $o(3)$
\begin{eqnarray}
\mathcal{L} \sim {e}(3)\oplus {o}(3) = \{\vec{J}-\vec{S}, \,\,\vec{\Pi}\}
\oplus \{\vec{S}\}
\,.\label{algdirectsum}
\end{eqnarray}

\section{Determining equations for an integral of motion}
\label{determineq}
\subsection{Derivation}
In this Subsection we give the full set of determining equations 
obtained from the commutativity condition $[H, X]=0$, where $H$ is the 
Hamiltonian given in (\ref{generalhamiltonianwithhsec1}) and $X$ is the 
most general first-order integral of motion given in 
(\ref{appendixmostgeneralintofmotion}). This commutator has 
second-, first- and zero-order terms in the momenta. By setting the 
coefficients of different powers of the momenta equal to zero in 
each entry of this $2 \times 2$ matrix we obtain the 
following determining equations. Since, the Planck constant 
$\hbar$ enters into the determiming equations in a nontrivial way we keep 
it throughout the whole set of determining equations. However, after 
giving the determining equations we set $\hbar=1$ for simplicity. 

\noindent
{\bf{i) Determining equations coming from the second-order terms}}

\noindent
From the diagonal elements it is immediately found that 
$A_0$, $B_0$ and $C_0$ are linear functions and are expressed 
for any potentials $V_0$ and $V_1$ as 
\begin{eqnarray}
A_0=b_1-a_3 y+a_2 z \,, \qquad 
B_0=b_2+a_3 x-a_1 z \,, \qquad
C_0=b_3-a_2 x+a_1 y \,, 
\label{A0B0C0}
\end{eqnarray}
where $a_i$ and $b_i$ ($i=1,2,3$) are real constants. After introducing 
(\ref{A0B0C0}) into the rest of the coefficients of the second-order terms 
and separating the imaginary and real parts of the coefficients coming from 
the off-diagonal elements we are left with an overdetermined system of 
eighteen partial differential equations for $A_i$, $B_i$, $C_i$ ($i=1,2,3$) 
and $V_1$. These are,
\begin{eqnarray}
&2zA_1V_1+\hbar A_{3,x}=0\,, \qquad 
2yA_1V_1+\hbar A_{2,x}=0\,, \qquad
2xB_2V_1+\hbar B_{1,y}=0\,, \qquad
2zB_2V_1+\hbar B_{3,y}=0\,, \nonumber \\
&2xC_3V_1+\hbar C_{1,z}=0\,, \qquad
2yC_3V_1+\hbar C_{2,z}=0\,, \qquad
2V_1\big(yA_2+zA_3\big)-\hbar A_{1,x}=0\,, \nonumber \\
&2V_1\big(xB_1+zB_3\big)-\hbar B_{2,y}=0\,, \qquad
2V_1\big(xC_1+yC_2\big)-\hbar C_{3,z}=0\,, \nonumber \\
&2zV_1\big(A_2+B_1\big)+\hbar A_{3,y}+\hbar B_{3,x}=0\,, \qquad
2yV_1\big(A_3+C_1\big)+\hbar A_{2,z}+\hbar C_{2,x}=0\,, \nonumber \\
&2xV_1\big(B_3+C_2\big)+\hbar B_{1,z}+\hbar C_{1,y}=0\,,  \qquad
2V_1\big(xA_1+yA_2-zC_1\big)-\hbar A_{3,z}-\hbar C_{3,x}=0\,, \nonumber \\
&2V_1\big(xB_1+yB_2-zC_2\big)-\hbar B_{3,z}-\hbar C_{3,y}=0\,, \qquad
2V_1\big(xA_2-yB_2-zB_3\big)+\hbar A_{1,y}+\hbar B_{1,x}=0\,, \nonumber \\
&2V_1\big(xA_1+zA_3-yB_1\big)-\hbar A_{2,y}-\hbar B_{2,x}=0\,, \qquad 
2V_1\big(xA_3-yC_2-zC_3\big)+\hbar A_{1,z}+\hbar C_{1,x}=0\,, \nonumber \\
&2V_1\big(yB_3-xC_1-zC_3\big)+\hbar B_{2,z}+\hbar C_{2,y}=0\,. \label{appendixsetof18eqs18}
\end{eqnarray}

\noindent
{\bf{ii) Determining equations coming from the first-order terms}}

\noindent
After introducing (\ref{A0B0C0}) and separating the real and imaginary 
parts, we have the following twelve partial differential equations
\begin{eqnarray}
&V_1\Big(\hbar (b_1-a_3 y) +2 y \phi_3\Big)+\hbar \Big(x(A_0 V_{1,x}+B_0 V_{1,y}+C_0 V_{1,z})+\phi_{2,z}\Big)=0\,, \nonumber \\
&V_1\Big(\hbar (b_1+a_2 z) -2 z \phi_2\Big)+\hbar \Big(x(A_0 V_{1,x}+B_0 V_{1,y}+C_0 V_{1,z})-\phi_{3,y}\Big)=0\,, \nonumber \\
&V_1\Big(\hbar (b_2-a_1 z) +2 z \phi_1\Big)+\hbar \Big(y(A_0 V_{1,x}+B_0 V_{1,y}+C_0 V_{1,z})+\phi_{3,x}\Big)=0\,, \nonumber \\
&V_1\Big(\hbar (b_2+a_3 x) -2 x \phi_3\Big)+\hbar \Big(y(A_0 V_{1,x}+B_0 V_{1,y}+C_0 V_{1,z})-\phi_{1,z}\Big)=0\,,  \nonumber \\
&V_1\Big(\hbar (b_3-a_2 x) +2 x \phi_2\Big)+\hbar \Big(z(A_0 V_{1,x}+B_0 V_{1,y}+C_0 V_{1,z})+\phi_{1,y}\Big)=0\,, \nonumber \\
&V_1\Big(\hbar (b_3+a_1 y) -2 y \phi_1\Big)+\hbar \Big(z(A_0 V_{1,x}+B_0 V_{1,y}+C_0 V_{1,z})-\phi_{2,x}\Big)=0\,, \nonumber \\
&V_1\Big(\hbar (a_2 y+a_3 z) -2 y \phi_2- 2 z \phi_3\Big)+\hbar \phi_{1,x}=0\,, \nonumber \\
&V_1\Big(\hbar (a_1 x+a_3 z) -2 x \phi_1- 2 z \phi_3\Big)+\hbar \phi_{2,y}=0\,, \nonumber \\
&V_1\Big(\hbar (a_1 x+a_2 y) -2 x \phi_1- 2 y \phi_2\Big)+\hbar \phi_{3,z}=0\,, \label{appendixcoeffirstord9}
\end{eqnarray}
\begin{eqnarray}
\phi_{0,x}&=&V_1\Big((yA_{3,x}-xA_{3,y})+(xA_{2,z}-zA_{2,x})+(zA_{1,y}-yA_{1,z})+(C_2-B_3)\Big) \nonumber \\
&\,&+V_{1,x}(zA_2-yA_3)+V_{1,y}(zB_2-yB_3)+V_{1,z}(zC_2-yC_3)\,,  \nonumber  \\
\phi_{0,y}&=&V_1\Big((yB_{3,x}-xB_{3,y})+(xB_{2,z}-zB_{2,x})+(zB_{1,y}-yB_{1,z})+(A_3-C_1)\Big) \nonumber \\
&\,&+V_{1,x}(xA_3-zA_1)+V_{1,y}(xB_3-zB_1)+V_{1,z}(xC_3-zC_1)\,,  \nonumber \\
\phi_{0,z}&=& V_1\Big((yC_{3,x}-xC_{3,y})+(xC_{2,z}-zC_{2,x})+(zC_{1,y}-yC_{1,z})+(B_1-A_2)\Big) \nonumber \\
&\,&+V_{1,x}(yA_1-xA_2)+V_{1,y}(yB_1-xB_2)+V_{1,z}(yC_1-xC_2)\,, \label{appendixcoeffirstordgen12} 
\end{eqnarray}
where $A_0$, $B_0$ and $C_0$ are given in (\ref{A0B0C0}). There are also 
nine other second-order partial differential equations 
for $A_i$, $B_i$, $C_i$ and $V_1$, coming from 
the coefficients of the first-order terms. However, these are differential 
consequences of (\ref{appendixsetof18eqs18}) so we do not present them here. 

\noindent
{\bf{iii) Determining equations coming from the zero-order terms}}

\noindent
Setting the coefficients of the zero-order terms in each entry of the 
commutation relation equal to zero and separating the real and imaginary 
parts, we have the following four partial differential equations
\begin{eqnarray}
A_0 V_{0,x}+B_0 V_{0,y}+C_0 V_{0,z}+V_1\big(x(\phi_{2,z}-\phi_{3,y})+y(\phi_{3,x}-\phi_{1,z})+z(\phi_{1,y}-\phi_{2,x})\big)=0\,, \nonumber 
\end{eqnarray}
\begin{eqnarray}
&\hbar \Big\{V_1\big(A_{1,z}+B_{2,z}-C_{2,y}-C_{1,x}\big)+V_{1,x}\big((xA_{1,z}-zA_{1,x})+(yA_{2,z}-zA_{2,y})\big) \nonumber \\
&+V_{1,y}\big((xB_{1,z}-zB_{1,x})+(yB_{2,z}-zB_{2,y})\big)+V_{1,z}\big((xC_{1,z}-zC_{1,x})+(yC_{2,z}-zC_{2,y})\big)\Big\} \nonumber \\
&+2\Big(A_3V_{0,x}+B_3V_{0,y}+C_3V_{0,z}\Big)+2V_1\Big(y\phi_{0,x}-x\phi_{0,y}\Big)=0\,,  \nonumber 
\end{eqnarray}
\begin{eqnarray}
&\hbar \Big\{V_1\big(B_{2,x}+C_{3,x}-A_{3,z}-A_{2,y}\big)+V_{1,x}\big((zA_{3,x}-xA_{3,z})+(yA_{2,x}-xA_{2,y})\big) \nonumber \\
&+V_{1,y}\big((zB_{3,x}-xB_{3,z})+(yB_{2,x}-xB_{2,y})\big)+V_{1,z}\big((zC_{3,x}-xC_{3,z})+(yC_{2,x}-xC_{2,y})\big)\Big\} \nonumber \\
&+2\Big(A_1V_{0,x}+B_1V_{0,y}+C_1V_{0,z}\Big)+2V_1\Big(z\phi_{0,y}-y\phi_{0,z}\Big)=0\,,  \nonumber 
\end{eqnarray}
\begin{eqnarray}
&\hbar \Big\{V_1\big(A_{1,y}+C_{3,y}-B_{3,z}-B_{1,x}\big)+V_{1,x}\big((zA_{3,y}-yA_{3,z})+(xA_{1,y}-yA_{1,x})\big) \nonumber \\
&+V_{1,y}\big((zB_{3,y}-yB_{3,z})+(xB_{1,y}-yB_{1,x})\big)+V_{1,z}\big((zC_{3,y}-yC_{3,z})+(xC_{1,y}-yC_{1,x})\big)\Big\} \nonumber \\
&+2\Big(A_2V_{0,x}+B_2V_{0,y}+C_2V_{0,z}\Big)+2V_1\Big(x\phi_{0,z}-z\phi_{0,x}\Big)=0\,. \label{appendixsecondrderphigeneral4}
\end{eqnarray}

In general the partial differential equations in (\ref{appendixsecondrderphigeneral4}) involve 
second- and third-order derivatives of $A_i$, $B_i$ and $C_i$, ($i=1,2,3$), 
however, using (\ref{appendixsetof18eqs18}) these terms can be eliminated.

\subsection{Discussion of solution in general case}
In general the solution of the determining equations 
(\ref{appendixsetof18eqs18})-(\ref{appendixsecondrderphigeneral4}) 
for the $15$ unknowns $\phi_0$, $V_0$, $V_1$ and $A_i$, $B_i$, $C_i$, 
$\phi_i$ ($i=1,2,3$) turns out to be a difficult problem. However, 
it is seen that the determining equations (\ref{appendixcoeffirstord9}) do 
not involve $\phi_0$, $A_i$, $B_i$, $C_i$ ($i=1,2,3$) and $V_0$ 
and hence could be analyzed separately. 

In order to determine the unknown functions $V_1$ and $\phi_i$, 
we express the first-order derivatives of $\phi_i$'s from (\ref{appendixcoeffirstord9}) and 
require the compatibility of the mixed 
partial derivatives. This requirement gives us another $9$ equations for 
$\phi_i$'s and first-order derivatives of them. Now, if we introduce the 
first-order derivatives of $\phi_i$'s from (\ref{appendixcoeffirstord9}) into this system, 
we get a system of algebraic equations for $\phi_i$'s ($i=1,2,3$). This 
system of algebraic equations can be written in the following way:
\begin{eqnarray}
 M.\Phi=R \label{symbolicalgebraicsystem}
\end{eqnarray}
where $M$ is a $9 \times 3$ matrix and $\Phi$ and $R$ are $3 \times 1$ 
and $9 \times 1$ vectors, respectively. 
The matrix $M$ can be written as:
\begin{eqnarray}
M=\left(\begin{array}{ccc}
0 & 2\delta_1 & 2z\delta_3 \\
0 & -2x\delta_2 & 2x\delta_3 \\
0 & 2y\delta_2 & 2\delta_5 \\
-2\delta_1 & 0 & -2z\delta_4 \\
2x\delta_2 & 0 & 2\delta_6 \\
-2y\delta_2 & 0  & 2y\delta_4 \\
-2z\delta_3 & 2z\delta_4  & 0  \\
-2x\delta_3 & -2\delta_6 & 0 \\
-2\delta_5 & -2y\delta_4 & 0 
\end{array}\right) \label{coefmatrixforphis}
\end{eqnarray}
where $\delta_i$ ($i=1,\ldots ,6$) are defined as follows
\begin{eqnarray}
&&\delta_1=2V_1-2z^2V_1^2+yV_{1,y}+xV_{1,x}\,, \qquad
\delta_2=2zV_1^2+V_{1,z}\,, \qquad
\delta_3=2yV_1^2+V_{1,y}\,, \nonumber \\
&&\delta_4=2xV_1^2+V_{1,x}\,, \qquad
\delta_5=2V_1-2y^2V_1^2+zV_{1,z}+xV_{1,x}\,, \qquad
\delta_6=2V_1-2x^2V_1^2+yV_{1,y}+zV_{1,z}\,. 
\label{elementscoefmatrixforphis}
\end{eqnarray}
The vector $\Phi$ is $(\phi_1, \phi_2, \phi_3)^T$ and the entries of the vector $R$ are given as follows:
\begin{eqnarray}
R_{11}=&\!\!\!\!2V_1\big(a_2-2 xz (A_0V_{1,x}+B_0V_{1,y}+C_0V_{1,z})\big)+a_2(xV_{1,x}+yV_{1,y}+zV_{1,z}) \nonumber \\
&\,-2(b_3 x+z A_0)V_1^2-b_3V_{1,x}-z\big(A_0V_{1,xx}+B_0V_{1,xy}+C_0V_{1,xz})\,, \nonumber \\
R_{21}=&\!\!\!\!y(A_0V_{1,xy}+B_0V_{1,yy}+C_0V_{1,yz})+z(A_0V_{1,xz}+B_0V_{1,yz}+C_0V_{1,zz}) \nonumber \\
&\,-2x(b_1+A_0)V_1^2+(3-4x^2V_1)(A_0V_{1,x}+B_0V_{1,y}+C_0V_{1,z})-b_1V_{1,x}, \nonumber \\
R_{31}=&\!\!\!\!2V_1\big(a_3+2 xy (A_0V_{1,x}+B_0V_{1,y}+C_0V_{1,z})\big)+a_3(xV_{1,x}+yV_{1,y}+zV_{1,z}) \nonumber \\
&\,+2(b_2 x+y A_0)V_1^2+b_2V_{1,x}+y\big(A_0V_{1,xx}+B_0V_{1,xy}+C_0V_{1,xz})\,, \nonumber \\
R_{41}=&\!\!\!\!-2V_1\big(a_1+2 yz (A_0V_{1,x}+B_0V_{1,y}+C_0V_{1,z})\big)-a_1(xV_{1,x}+yV_{1,y}+zV_{1,z}) \nonumber \\
&\,-2(b_3 y+z B_0)V_1^2-b_3V_{1,y}-z\big(A_0V_{1,xy}+B_0V_{1,yy}+C_0V_{1,yz})\,, \nonumber \\
R_{51}=&\!\!\!\!2V_1\big(a_3-2 xy (A_0V_{1,x}+B_0V_{1,y}+C_0V_{1,z})\big)+a_3(xV_{1,x}+yV_{1,y}+zV_{1,z}) \nonumber \\
&\,-2(b_1 y+x B_0)V_1^2-b_1V_{1,y}-x\big(A_0V_{1,xy}+B_0V_{1,yy}+C_0V_{1,yz})\,, \nonumber \\
R_{61}=&\!\!\!\!-x(A_0V_{1,xx}+B_0V_{1,xy}+C_0V_{1,xz})-z(A_0V_{1,xz}+B_0V_{1,yz}+C_0V_{1,zz}) \nonumber \\
&\,+2y(b_2+B_0)V_1^2-(3-4y^2V_1)(A_0V_{1,x}+B_0V_{1,y}+C_0V_{1,z})+b_2V_{1,y}, \nonumber \\
R_{71}=&\!\!\!\!x(A_0V_{1,xx}+B_0V_{1,xy}+C_0V_{1,xz})+y(A_0V_{1,xy}+B_0V_{1,yy}+C_0V_{1,yz}) \nonumber \\
&\,-2z(b_3+C_0)V_1^2+(3-4z^2V_1)(A_0V_{1,x}+B_0V_{1,y}+C_0V_{1,z})-b_3V_{1,z}, \nonumber \\
R_{81}=&\!\!\!\!-2V_1\big(a_2+2 xz (A_0V_{1,x}+B_0V_{1,y}+C_0V_{1,z})\big)-a_2(xV_{1,x}+yV_{1,y}+zV_{1,z}) \nonumber \\
&\,-2(b_1 z+x C_0)V_1^2-b_1V_{1,z}-x\big(A_0V_{1,xz}+B_0V_{1,yz}+C_0V_{1,zz})\,, \nonumber \\
R_{91}=&\!\!\!\!2V_1\big(-a_1+2 yz (A_0V_{1,x}+B_0V_{1,y}+C_0V_{1,z})\big)-a_1(xV_{1,x}+yV_{1,y}+zV_{1,z}) \nonumber \\
&\,+2(b_2 z+y C_0)V_1^2+b_2V_{1,z}+y\big(A_0V_{1,xz}+B_0V_{1,yz}+C_0V_{1,zz})\,,
\end{eqnarray}
where $A_0$, $B_0$ and $C_0$ are given in (\ref{A0B0C0}).

In general the rank of matrix $M$ is $3$ and the rank of the extended 
matrix of the system (\ref{symbolicalgebraicsystem}) is $4$. Thus, we have to analyze under which conditions we can 
equate these two ranks. The rank of matrix $M$ can be $0$, $2$ or $3$. 
It can be $2$ if and only if $V_1$ satisfies the 
following differential equation
\begin{equation}
 2V_1+xV_{1,x}+yV_{1,y}+zV_{1,z}=0\,,
\end{equation}
which implies 
\begin{equation}
V_1=\frac{F(\xi, \eta)}{r^2}\,, \qquad F\ne 1\,,
\end{equation}
where $\xi=\frac{x}{z}$ and $\eta=\frac{y}{z}$. It can be $0$ if and only if 
$V_1=\frac{1}{r^2}$. Hence, the rank of matrix $M$ and the rank of the 
extended matrix can be equal if we have the folowing 
\begin{enumerate}
 \item[i)] For rank$=0$, $V_1=\frac{1}{r^2}$ with $a_i\neq0$ and $b_i\neq0$, $i=1,2,3$ 
 \item[ii)] For rank$=3$, $V_1=V_1(r)$ and $b_i=0$ for all $i$ and $a_i\neq0$, $i=1,2,3$.
\end{enumerate}

For the former case the system (\ref{symbolicalgebraicsystem}) has 
the following solution
\begin{eqnarray}
&\,&\phi_1=\frac{a_1}{2}+\frac{(x^2-y^2-z^2)\alpha_1+2xy\alpha_2+2xz\alpha_3-2b_2z+2b_3y}{2(x^2+y^2+z^2)}\,, \nonumber \\
\,\nonumber \\
&\,&\phi_2=\frac{a_2}{2}+\frac{(y^2-x^2-z^2)\alpha_2+2xy\alpha_1+2yz\alpha_3+2b_1z-2b_3x}{2(x^2+y^2+z^2)}\,, \nonumber \\
\,\nonumber \\
&\,&\phi_3=\frac{a_3}{2}+\frac{(z^2-x^2-y^2)\alpha_3+2xz\alpha_1+2yz\alpha_2-2b_1y+2b_2x}{2(x^2+y^2+z^2)}\,, \label{sonphi3}
\end{eqnarray}
where $\alpha_1$, $\alpha_2$ and $\alpha_3$ are real constants and for the latter it has 
\begin{eqnarray}
\phi_1=\frac{a_1}{2}\,, \qquad \phi_2=\frac{a_2}{2}\,, \qquad \phi_3=\frac{a_3}{2}\,. \label{phisforspherisymmetricV1}
\end{eqnarray}

Indeed for the potential $V_1=\frac{1}{r^2}$, the whole set of determining 
equations 
(\ref{appendixsetof18eqs18})-(\ref{appendixsecondrderphigeneral4}) 
can be solved and we obtain a $9$-dimensional Lie algebra $\mathcal{L}$ 
given in (\ref{algdirectsum}). For $V_1=V_1(r)$, $V_0=V_0(r)$ we obtain the 
well-known result that $H$ commutes with total angular momentum 
$\vec{J}=\vec{L}+\frac{1}{2}\vec{\sigma}$.

In the rest of the article we restrict to spherically symmetric potentials 
$V_1=V_1(r)$, $V_0=V_0(r)$ and have a rotationally invariant Hamiltonian. 
However, the integrals of motion would transform under the action of rotations 
$O(3)$. Instead of solving the whole set 
of determining equations we analyze the problem by classifying the 
integrals of motion into irreducible $O(3)$ multiplets.

\section{Classification of integrals of motion 
into $O(3)$ multiplets}\label{o3multiplets}
Let us assume that the Hamiltonian is 
\begin{eqnarray}
H=-\frac{1}{2} \Delta + V_0(r) + V_1(r)\, \vec{\sigma}\cdot \vec{L}\,,
\label{generalhamiltonianwithh}
\end{eqnarray}
and that $X$ of (\ref{appendixmostgeneralintofmotion}) 
is an integral of motion. Rotations in $E(3)$ leave 
the Hamiltonian invariant but can 
transform $X$ into new invariants. We can hence decompose the space 
of integrals of motion into subspaces transforming under irreducible 
representations of $O(3)$. We shall also require that the subspaces 
have definite behavior under the parity operator. 

At our disposal are two vectors $\vec{x}$ and $\vec{p}$ and one 
pseudovector $\vec{\sigma}$. The integrals we are considering 
can involve at most first-order powers of $\vec{p}$ and $\vec{\sigma}$ 
but arbitrary powers of $\vec{x}$.

General formulas for the decomposition of the representation 
$[D(j)]^n$ of $O(3)$ into irreducible components are given by 
Murnaghan \cite{Murnaghan}. Since we are interested only in 
$j=1$ we shall proceed ab initio rather than specialize his 
results. 

We shall construct scalars, pseudo-scalars, vectors, axial vectors and 
symmetric two component tensors and pseudotensors in the space
\begin{eqnarray}
\Big\{\{\vec{x}\}^n \times \vec{p} \times \vec{\sigma}\Big\}\,.
\label{generalspace}
\end{eqnarray}

The quantities $\vec{x}$, $\vec{p}$ and $\vec{\sigma}$ allow us to 
define $6$ independent ``directions'' in the direct product of the 
Euclidean space and the spin one, namely 
\begin{eqnarray}
\Big\{\vec{x},\, \vec{p},\, \vec{L}=\vec{x}\wedge\vec{p},\, \vec{\sigma},\, \vec{\sigma}\wedge\vec{x},\, \vec{\sigma}\wedge\vec{p}\Big\}\,,
\label{sixdimspace}
\end{eqnarray}
and any $O(3)$ tensor can be expressed in terms of these. The positive integer $n$ in (\ref{generalspace}) is arbitrary and any scalar in $\vec{x}$ space will be 
written as $f(r)$ where $f$ is an arbitrary function of 
$r=\sqrt{x^2+y^2+z^2}$. Since $\vec{\sigma}$ and $\vec{p}$ figure at most 
linearly we can form exactly $3$ independent scalars and $3$ pseudoscalars 
out of the quantities (\ref{sixdimspace}): 

Scalars
\begin{eqnarray}
S_1=1\,, \qquad S_2=(\vec{x}, \vec{p})\,, \qquad S_3=(\vec{\sigma}, \vec{L})\,.
\label{scalars}
\end{eqnarray}

Pseudoscalars
\begin{eqnarray}
P_1=(\vec{\sigma}, \vec{x})\,, \qquad P_2=(\vec{\sigma}, \vec{p})\,, 
\qquad P_3=(\vec{x}, \vec{p})(\vec{\sigma}, \vec{x})\,.
\label{pseudoscalars}
\end{eqnarray}

The independent vectors and axial vectors are:

Vectors
\begin{eqnarray}
&&\vec{V}_1=\vec{x}\,, \qquad \vec{V}_2=\vec{p}\,, \qquad \vec{V}_3=(\vec{x}, \vec{p})\,\vec{x}\,, 
\qquad \vec{V}_4=(\vec{\sigma}, \vec{L})\,\vec{x}\,, \qquad \vec{V}_5=(\vec{x}, \vec{p})(\vec{\sigma}\wedge\vec{x})\,, \nonumber \\
&&\vec{V}_6= \vec{\sigma}\wedge\vec{p}\,, \qquad \vec{V}_7=\vec{\sigma}\wedge\vec{x}\,, \qquad \vec{V}_8=(\vec{\sigma}, \vec{x})\,\vec{L}\,.
\label{vectors}
\end{eqnarray}

Axial vectors
\begin{eqnarray}
&&\vec{A}_1=\vec{L}\,, \qquad \vec{A}_2=\vec{\sigma}\,, \qquad \vec{A}_3=(\vec{x}, \vec{p})\,\vec{\sigma}\,, 
\qquad \vec{A}_4=(\vec{\sigma}, \vec{p})\,\vec{x}\,, \qquad \vec{A}_5=(\vec{x}, \vec{\sigma})\,\vec{x}\,, \nonumber \\
&&\vec{A}_6= (\vec{x}, \vec{\sigma})\,\vec{p}\,, \qquad \vec{A}_7=(\vec{x}, \vec{p})(\vec{\sigma}, \vec{x})\,\vec{x}\,.
\label{axialvectors}
\end{eqnarray}

Similarly we can form $10$ independent $2$-component symmetric tensors and 
$9$ symmetric pseudotensors:

Tensors
\begin{eqnarray}
&&T_1^{ik}=x^i x^k\,, \qquad T_2^{ik}=(\vec{x}, \vec{p})\,x^i x^k\,, \qquad T_3^{ik}=(\vec{\sigma}, \vec{L})\,x^i x^k\,, \qquad T_4^{ik}=x^i p^k + x^k p^i\,, \nonumber \\
&&T_5^{ik}=(\vec{\sigma}, \vec{x})\,(x^i L^k +x^k L^i)\,, \qquad T_6^{ik}=x^i (\vec{\sigma}\wedge\vec{x})^k + x^k (\vec{\sigma}\wedge\vec{x})^i\,, \nonumber \\
&&T_7^{ik}=(\vec{x}, \vec{p})(x^i (\vec{\sigma}\wedge\vec{x})^k + x^k (\vec{\sigma}\wedge\vec{x})^i)\,, \qquad 
T_8^{ik}=x^i (\vec{\sigma}\wedge\vec{p})^k + x^k (\vec{\sigma}\wedge\vec{p})^i\,, \nonumber \\
&&T_9^{ik}=p^i (\vec{\sigma}\wedge\vec{x})^k + p^k (\vec{\sigma}\wedge\vec{x})^i\,, \qquad T_{10}^{ik}=L^i\sigma^k + L^k \sigma^i\,.
\label{tensors}
\end{eqnarray}

Pseudotensors
\begin{eqnarray}
&&Y_1^{ik}=(\vec{\sigma}, \vec{p})\,x^i x^k\,, \quad Y_2^{ik}=(\vec{\sigma}, \vec{x})\,x^i x^k\,, \quad Y_3^{ik}=(\vec{x}, \vec{p})(\vec{\sigma}, \vec{x})\,x^i x^k\,, \nonumber \\
&&Y_4^{ik}=(\vec{\sigma}, \vec{x})\,(x^i p^k + x^k p^i)\,, \quad Y_5^{ik}=x^i L^k +x^k L^i\,, \quad Y_6^{ik}=x^i \sigma^k + x^k \sigma^i\,, \nonumber \\
&&Y_7^{ik}=(\vec{x}, \vec{p})\,(x^i \sigma^k + x^k \sigma^i)\,, \quad 
Y_8^{ik}=p^i \sigma^k + p^k \sigma^i\,, \quad Y_9^{ik}=L^i (\vec{\sigma}\wedge\vec{x})^k + L^k (\vec{\sigma}\wedge\vec{x})^i\,.
\label{pseudotensors}
\end{eqnarray}

An arbitrary function $f(r)$ is also a scalar and each of the 
quantities in (\ref{scalars})-(\ref{pseudotensors}) can be 
multiplied by $f(r)$ without changing its properties under 
rotations or reflections. 

All of the tensors and pseudotensors should be considered to be 
traceless since their traces appear separately as scalars, or 
pseudoscalars. For simplicity of notation we do not subtract the 
trace explicitly.

\section{Solution of Commutativity Equations for 
$V_0=V_0(r)$, $V_1=V_1(r)$}\label{solutionsofcommeq}
In this Section we separately take the linear combinations of all 
the scalars, pseudo-scalars, vectors, axial-vectors and 
two component tensors and pseudotensors. For simplicity of notation 
we just write the bare linear combinations, however, in the analysis 
of the commutation relations we use the full symmetric form of those, 
which could be found in the Appendix.

\subsection{Scalars}
Let us take a linear combination of the scalars given in 
(\ref{scalars})
\begin{equation}
X_S=\sum_{j=1}^3 f_j(r) S_j\,. 
\label{linscalars}
\end{equation}
It is immediately seen that in order to satisfy the 
commutativity equation $[H, X_S]=0$ we must have 
\begin{equation}
f_1=c_1\,, \qquad f_2=0\,, \qquad f_3=c_2\,,
\label{sollinscalars}
\end{equation}
where $c_1$ and $c_2$ are real constants. The corresponding 
integrals $S_1$ and $S_3$ are trivial. 


\subsection{Pseudoscalars}
As an integral of motion we take a linear combination of 
the pseudoscalars given in (\ref{pseudoscalars})
\begin{equation}
X_P=\sum_{j=1}^3 f_j(r) P_j\,,  
\label{linpseudoscalars}
\end{equation}
and require $[H, X_P]=0$. The determining equations, obtained by 
equating the coefficients of the second-order terms to zero in the commutativity 
equation, become 
\begin{eqnarray}
f_3^{\prime} = -2 r f_3 V_1\,, \qquad f_3 = -2 f_2 V_1\,, \label{secorcombapseudoscalarone} \\
2 r V_1^2\big(2r^2V_1-3\big)-V_1^{\prime} = 0\,.
\label{secorcombapseudoscalar}
\end{eqnarray}

Depending on the solutions of the compatibility equation for $V_1$ 
(\ref{secorcombapseudoscalar}) we have several cases. The solution of 
(\ref{secorcombapseudoscalar}) is given by 
\begin{equation}
V_1=\frac{1}{2r^2} \left(1+\frac{\alpha}{\sqrt{1+\beta r^2}}\right)\,,
\label{V1pseudoscalar}
\end{equation}
where $\alpha^2=1$. Note that $V_1=\frac{1}{r^2}$ and $V_1=\frac{1}{2r^2}$ 
are special solutions of (\ref{secorcombapseudoscalar}) with 
$(\alpha, \beta) = (1,0)$ and $(1, \infty)$, respectively. The case 
$V_1=\frac{1}{r^2}$ induced by a gauge transformation has already been 
considered. 

{\bf Case I:} $V_1=\frac{1}{2r^2}$

For this type of potential, (\ref{secorcombapseudoscalarone}) implies that 
\begin{equation}
f_2 = -c_1 r\,, \qquad {\rm and} \qquad f_3=\frac{c_1}{r}\,, 
\end{equation}
and the first-order equations give 
\begin{equation}
f_1 =\frac{c_2}{r} \,. 
\end{equation}
Then zero-order equations are satisfied for any $V_0(r)$ and we have 
an integral of motion $X_P$ for the above values of $f_i$.

{\bf Case II:} $V_1=\frac{1}{2r^2} 
\left(1+\frac{\alpha}{\sqrt{1+\beta r^2}}\right)\,, \qquad 0< \beta < \infty\,, \qquad \alpha^2=1$

For this type of potential, (\ref{secorcombapseudoscalarone}) implies that 
\begin{equation}
f_2 = -\frac{c_1}{\beta}\sqrt{1+\beta r^2}\,, \qquad {\rm and} \qquad f_3=\frac{c_1}{-\alpha +\sqrt{1+\beta r^2}}\,, 
\end{equation}
and the first-order equations give $f_1 = 0$. 
Then $V_0$ is determined from the zero-order equatiosns to be $V_0=V_1$ 
and we have an integral of motion $X_P$ for the above values of $f_i$.

\subsection{Vectors}
Let us now take the linear combination of the vectors given in 
(\ref{vectors})
\begin{equation}
\vec{X}_V=\sum_{j=1}^8 f_j(r) \vec{V}_j\,,  
\label{linvectors}
\end{equation}
and require $[H, \vec{X}_V]=0$. The second-order terms give 
\begin{eqnarray}
&&f_2 = c_1\,, \qquad f_3=0\,, \qquad f_4 + f_5 =0\,,  \\
&&r^2 f_4^{\prime} = f_6^{\prime} -2r f_4\,, \label{two} \\
&&f_4-2f_6 V_1 +f_8 (1-2r^2 V_1) = 0\,, \label{secordvectors} \\
&&f_8^{\prime} = -2rV_1(f_8 +f_4)-f_4^{\prime}\,.\label{one} 
\end{eqnarray}
Equation (\ref{two}) implies 
\begin{equation}
f_4=\frac{c_2 +f_6}{r^2}\,.
\label{solf4}
\end{equation}
Introducing (\ref{solf4}) into (\ref{secordvectors}) and solving for 
$f_8$ we obtain 
\begin{equation}
f_8=-\frac{c_2 + f_6 (1-2r^2 V_1)}{r^2(1-2r^2V_1)}\,.
\label{solf8}
\end{equation}
Finally if we introduce $f_4$ and $f_8$ into (\ref{one}) we get a 
compatibility condition for $V_1$ which is exactly same as 
(\ref{secorcombapseudoscalar}). Thus we have the following cases: 

{\bf Case I:} $V_1=\frac{1}{2r^2}$

For this type of potential, the commutativity equation $[H, \vec{X}_V]=0$ 
implies $f_2 = f_3 = f_7 = 0$,  $f_4 = -f_5 = \frac{f_6}{r^2}, f_1 = \frac{c_3}{2r}$ and $f_8 = \frac{c_3}{r} - \frac{f_6}{r^2}$. Thus we have an integral of motion $\vec{X}_V$ for these values of $f_i$.

{\bf Case II:} $V_1=\frac{1}{2r^2} \left(1+\frac{\alpha}{\sqrt{1+\beta r^2}}\right)$

For this type of potential, the commutativity equation $[H, \vec{X}_V]=0$ 
implies $f_1 = f_2 = f_3 = f_7 = 0$,  $-f_4 = f_5 = f_8 = -\frac{f_6}{r^2}$. However, if we introduce these values of $f_i$ into the integral of motion $\vec{X}_V$ in (5.10), it identically vanishes. Hence, we do not have an integral of motion.

{\bf Case III:} $V_1=V_1(r)$ and $f_4 =0\,, f_6=0$, and $f_8=0$

For this case equations (\ref{two})-(\ref{one}) are all 
satisfied and we have 
\begin{equation}
f_2=c_1\,, \quad f_3 = 0\,, \quad  f_4 = 0\,, \quad f_5 = 0\,, \quad f_6 = 0\,, \quad f_8 = 0\,.
\end{equation}
Then first-order terms determine $f_1$, $f_7$ as 
\begin{eqnarray}
f_1=0\,, \qquad f_7^{\prime} + 2rf_7 V_1 = 0\,, \qquad 2r f_7 V_1 + c_1 V_1^{\prime}=0\,,
\end{eqnarray}
and we conclude that we either have $V_1=\frac{1}{r^2}$ or we do not have 
any integral of motion (i.e. $f_j=0$ for all $j=1,\ldots,8$).

\subsection{Axial vectors}
Let us now take the linear combination of the axial vectors given in 
(\ref{axialvectors})
\begin{equation}
\vec{X}_A=\sum_{j=1}^7 f_j(r) \vec{A}_j\,,  
\label{linaxialvectors}
\end{equation}
and require $[H, \vec{X}_A]=0$. The second-order terms give 
\begin{eqnarray}
&&f_1=c_1\,, \qquad f_3=0\,, \label{axsec1} \\
&&f_4^{\prime} = 4rf_4V_1(1-r^2V_1)\,, \label{axsec2} \\ 
&&f_6^{\prime} = 2r(f_4-f_6) V_1\,, \qquad f_4 = -f_6 (1-2r^2V_1)\,, \label{axsec3} \\
&&f_7^{\prime} + 2rf_7V_1=0\,, \qquad f_7=-2f_4V_1\,. \label{axsec4} 
\end{eqnarray}
Equations (\ref{axsec2}) and (\ref{axsec3}) imply 
\begin{equation}
f_6\Big(2V_1(3-6r^2V_1 + 4r^4V_1^2) + rV_1^{\prime}\Big)=0\,,
\label{axf6}
\end{equation}
where as (\ref{axsec2}) and (\ref{axsec4}) imply 
\begin{equation}
f_4\Big(2rV_1^2(3-2r^2V_1) + V_1^{\prime}\Big)=0\,. 
\label{axf4}
\end{equation}
The only common solutions of (\ref{axf6}) and (\ref{axf4}) are 
$V_1=\frac{1}{r^2}$ and $V_1=\frac{1}{2r^2}$. Thus, other than 
$V_1=\frac{1}{r^2}$, we have two cases:


{\bf Case I:} $V_1=\frac{1}{2r^2}$

For this type of potential, (\ref{axsec2})-(\ref{axsec4}) give 
\begin{equation}
f_4=0\,, \qquad f_6=\frac{c_2}{r^2}\,, \qquad f_7=0\,,
\label{axcase2f4}
\end{equation}
and then the first-order terms imply 
\begin{eqnarray}
f_2=\frac{c_1}{2}\,, \qquad f_5=0\,, \qquad c_2=0\,,
\label{axcase2f2f5}
\end{eqnarray}
The zero-order terms are satisfied for arbitrary $V_0$. 
Thus the only integral of motion is 
$\vec{X}_A = \vec{L} + \frac{1}{2}\vec{\sigma}$, the total 
angular momentum (an integral for any $V_1(r)$ and $V_0(r)$).

{\bf Case II:} $V_1=V_1(r)$ and $f_4=0,\,f_6=0,\, f_7=0$

For $f_4=0,\,f_6=0,\, f_7=0$, (\ref{axsec2})-(\ref{axsec4}) are all 
satisfied for arbitrary $V_1(r)$. The first-order terms give 
\begin{eqnarray}
&&f_2=c_2\,, \qquad f_5=(c_1-2c_2)V_1\,, \label{axcase3f2f5} \\
&&f_5^{\prime} + 2r f_5 V_1 =0\,.
\label{axcase3comV1}
\end{eqnarray}
However, (\ref{axcase3comV1}) together with (\ref{axcase3f2f5}) imply 
\begin{eqnarray}
(c_1-2c_2)(V_1^{\prime} + 2r V_1^2) = 0\,.
\end{eqnarray}
Thus we have two more subcases: 

i) $c_1=2c_2$

Then we have $f_5=0$ and $f_2=\frac{c_1}{2}$ and the rest of the 
determining equations are satisfied for arbitrary $V_1(r)$ and 
$V_0(r)$. Hence the only integral of motion is the total angular 
momentum. 

ii) $V_1^{\prime} + 2r V_1^2 = 0$

We have $V_1=\frac{1}{r^2-\alpha}$. Introducing this $V_1$ 
together with (\ref{axcase3f2f5}) into the determining equations 
we obtain $(c_1-2c_2)\alpha=0$. Hence we conclude that we either have 
$\alpha=0$ (i.e. $V_1=\frac{1}{r^2}$) or $c_1=2c_2$, both of which 
have already been investigated. 


\subsection{Tensors}
Let us now take the linear combination of the tensors given in 
(\ref{tensors}):
\begin{equation}
X_T^{ik}=\sum_{j=1}^{10} f_j(r) {T}_j^{ik}\,,  
\label{lintensors}
\end{equation}
and require $[H, X_T^{ik}]=0$. The second-order terms give 
\begin{eqnarray}
&&f_2=0\,, \qquad f_4=0\,, \qquad 
f_3^{\prime} = -2 f_7^{\prime}\,, \label{secten2} \\ 
&&f_3 + f_5 + f_7 +2 (f_{10} -f_9) V_1 = 0\,, \qquad 
f_3 + 2f_5 +4r^2(f_{10} -f_9) V_1^2 = 0\,, \label{secten3} \\
&&f_7 - f_5 +2(r^2f_5 + f_8 + f_{10}) V_1 = 0\,, \label{secten5} \\
&&f_7^{\prime} -2r (f_5 - f_7) V_1 - f_5^{\prime} = 0\,, \label{secten6} \\
&&f_8 + f_9 + r^2\big(f_5 + 2(f_{10} -f_9) V_1 \big) = 0\,, \label{secten7} \\
&&f_{10}^{\prime} - f_9^{\prime} + 4r(f_{10} -f_9) V_1 (1-r^2V_1) = 0\,, \label{secten8} \\
&&(f_{10} -f_9)\Big(2V_1(3-6r^2V_1+4r^4V_1^2) + rV_1^{\prime} \Big) = 0\,.\label{secten9}
\end{eqnarray}
Equation (\ref{secten9}) implies either $f_{10} = f_9$ or a compatibility 
condition for $V_1$ which has the following solution
\begin{equation}
V_1 = \frac{1}{2r^2}\pm \frac{1}{\sqrt{4r^4 + \alpha}}\,. \label{tenV1}
\end{equation}
Notice that the potentials $V_1 = \frac{1}{r^2}$ and $V_1 = \frac{1}{2r^2}$ are 
also solutions which correspond to limiting values of $\alpha$ 
(i.e. $\alpha = 0$ and $\alpha = \infty$). Thus we have the following cases:

{\bf Case I:} $f_{10} = f_9$

Equations (\ref{secten3})-(\ref{secten7}) give 
\begin{eqnarray}
f_3 = -2f_5\,, \qquad f_7 = f_5\,, \qquad f_8 = -(f_{10} + r^2f_5)\,,
\end{eqnarray}
for arbitarry $V_1(r)$ and the first-order terms imply 
\begin{eqnarray}
f_1 = 0\,, \qquad f_6 = 0\,.
\end{eqnarray}
Then the zero-order terms are satisfied for any $V_1(r)$ and 
$V_0(r)$. However, if we introduce the above values of $f_i$ into the 
integral of motion $X_T^{12}$ in (\ref{lintensors}), it identically 
vanishes. Hence, we do not have any integral of motion for this case.


{\bf Case II:} $V_1 = \frac{1}{2r^2}$

For this type of potential, (\ref{secten3})-(\ref{secten8}) imply 
\begin{equation}
f_{10} - f_9 = \frac{c_1}{r}\,, \qquad f_3 +2f_5 = -\frac{c_1}{r^3}\,, \qquad r^2f_5 + f_8 + f_{10} =0\,, \qquad f_7 = f_5\,.
\end{equation}
However, then (\ref{secten2}) implies $c_1 = 0$ and we are back in Case I. 


{\bf Case III:} $V_1 = \frac{1}{2r^2}\pm \frac{1}{\sqrt{4r^4 + \alpha}}$

For this type of potential, (\ref{secten8}) implies 
\begin{equation}
f_{10} = \frac{c_1}{r}(4r^4 + \alpha)^{\frac{1}{4}} + f_9\,,
\end{equation}
and (\ref{secten5}) together with (\ref{secten7}) give 
\begin{equation}
f_7 = \pm \frac{4c_1r}{(4r^4 + \alpha)^{\frac{1}{4}}}V_1 + f_5\,. \label{tensorsf7}
\end{equation}
However, if we introduce (\ref{tensorsf7}) into (\ref{secten6}) we see that 
we must have 
\begin{equation}
c_1\alpha = 0\,.
\end{equation}
Hence, we either have $c_1 = 0$ or $\alpha = 0$, both of which have 
already been investigated. 


\subsection{Pseudotensors}
Let us now take the linear combination of the pseudotensors given in 
(\ref{pseudotensors})
\begin{equation}
X_Y^{ik}=\sum_{j=1}^9 f_j(r) {Y}_j^{ik}\,,  
\label{linpseudotensors}
\end{equation}
and require $[H, X_Y^{ik}]=0$. The second-order terms give 
\begin{eqnarray}
&&f_3=-2(f_1 + 2f_9) V_1\,, \quad f_4 = -2f_8 V_1 + (1-2r^2V_1)f_9\,, \quad f_5=0\,, \quad f_7 = f_9\,, \label{secpsten1} \\
&&f_1=2\left(2f_8V_1 (1-r^2V_1) - f_9 \left(1-2r^2 V_1 (1-r^2V_1) \right) \right)\,, \quad f_8^{\prime} = -r(2f_9 + r f_9^{\prime})\,, \label{secpsten2} \\
&&(f_8 + r^2 f_9)\big(2rV_1^2(3-2r^2V_1) - V_1^{\prime}\big) + f_9^{\prime} = 0\,, \label{secpsten3} \\
&&(f_8 + r^2 f_9)\Big(2rV_1^2(3-4r^2V_1 + 2r^4V_1^2) - (1-2r^2V_1) V_1^{\prime}\Big) = 0\,, \label{secpsten4} \\
&&(f_8 + r^2 f_9)\left(2rV_1^2(3-4r^2V_1) + \left( 6r^2V_1(1-r^2V_1) - 1\right) V_1^{\prime}\right) = 0\,. \label{secpsten5}
\end{eqnarray}
Equations (\ref{secpsten4}) and (\ref{secpsten5}) imply that we either have 
$f_8 + r^2 f_9 = 0$, or $V_1=\frac{1}{r^2}$. Hence, other than 
$V_1=\frac{1}{r^2}$, we have the following case:

{\bf Case I:} $f_8 + r^2 f_9 = 0$

Equations (\ref{secpsten1})-(\ref{secpsten3}) give 
\begin{eqnarray}
f_1 = 2c_1\,, \qquad f_3 = 0\,, \qquad f_4 = f_7 = f_9 = -c_1\,, \qquad f_8 = r^2 c_1\,.
\end{eqnarray}
Then the first-order terms give 
\begin{eqnarray}
f_2 = 0\,, \qquad f_6 = 0\,.
\end{eqnarray}
However, upon introducing the above values of $f_j$, ($j=1,\ldots,9$) 
into the integral of motion $X_Y^{12}$ in (\ref{linpseudotensors}), 
it vanishes identically and we do not have an integral of motion for 
this case. 


\section{Discussion of results}\label{results}
For any spherically symmetric potentials 
$V_0(r)$ and $V_1(r)$ it is well known that 
$J_i = L_i + \frac{1}{2} \sigma_i$ and $(\vec{\sigma}, \vec{L})$ 
are integrals of motion, where $\vec{J}$ is an axial vector 
and $(\vec{\sigma}, \vec{L})$ is a scalar. 

Additional first-order integrals exist only in four 
special cases. Two of them are treated in 
Sections \ref{subgt1} and \ref{subgt2} and 
the spin-orbital term $V_1 = \frac{1}{r^2}$ is 
gauge induced. 

For $V_0 = V_1 = \frac{1}{r^2}$ we 
obtain the vector $\vec{\Pi}$ and the axial vectors 
$\vec{J}$ and $\vec{S}$ of (\ref{generatorsforninedimalg}). 
In addition we obtain:

\noindent 
{\bf Pseudoscalar}
\begin{eqnarray}
X_P = -\frac{1}{2}\,(\vec{\sigma}, \vec{p}) + \frac{1}{r^2}\,(\vec{\sigma}, \vec{x})\,(\vec{x}, \vec{p}) - \frac{i}{r^2}\,(\vec{\sigma}, \vec{x})\,.
\end{eqnarray}

\noindent
{\bf Vector}
\begin{eqnarray}
\vec{V} = \frac{2\vec{x}}{r^2} - (\vec{\sigma} \wedge \vec{p}) + \frac{2}{r^2} (\vec{\sigma}, \vec{x})\,(\vec{x} \wedge \vec{p}) - \frac{i(\vec{x} \wedge \vec{\sigma})}{r^2}\,.
\end{eqnarray}

\noindent
{\bf Axial vector}
\begin{eqnarray}
\vec{A} = -\frac{1}{2}\,\vec{x}\,(\vec{\sigma}, \vec{p}) + \frac{1}{2}\vec{\sigma} - \frac{1}{2}\,(\vec{\sigma}, \vec{x})\,\vec{p} + \frac{\vec{x}}{r^2}\,(\vec{\sigma}, \vec{x})\,(\vec{x}, \vec{p}) -i \frac{3\vec{x}}{2r^2} (\vec{\sigma}, \vec{x})\,.
\end{eqnarray}
However, they all lie in the enveloping algebra of the Lie algebra 
(\ref{algdirectsum}). 

For $V_1 = \frac{1}{r^2}$ and $V_0 = V_0(r) \neq V_1$ we reobtain 
the algebra (\ref{algdirectsumo3o3}) and the gauge transforms 
of the axial vector $\vec{\sigma} \wedge \vec{L}$ and the 
tensor $\sigma^j L^k$, that are in the enveloping algebra of 
(\ref{algdirectsumo3o3}).  

For $V_1=\frac{1}{2r^2}$ and $V_0=V_0(r)$ we have:

\noindent
{\bf Pseudoscalars}
\begin{eqnarray}
&&X_P^1 = \frac{(\vec{\sigma}, \vec{x})}{r}\,, \label{solintofmotforpsone} \\
&&X_P^2 = -r (\vec{\sigma}, \vec{p}) + \frac{1}{r}\,(\vec{\sigma}, \vec{x})\,(\vec{x}, \vec{p}) - \frac{i}{r}\,(\vec{\sigma}, \vec{x})\,.
\label{solintofmotforpstwo}
\end{eqnarray}

\noindent
{\bf Vector}
\begin{eqnarray}
\vec{X}_V = \frac{1}{2r} \vec{x} + \frac{1}{r} (\vec{\sigma}, \vec{x}) \vec{L} - \frac{i}{2r} (\vec{x} \wedge \vec{\sigma})\,.
\label{solintofmotforvec}
\end{eqnarray}

For $V_0 = V_1 = \frac{1}{2r^2} \left(1+\frac{\alpha}{\sqrt{1+\beta r^2}}\right)$, $\alpha^2 = 1$ 
we have the following pseudoscalar

\begin{eqnarray}
X_P = -\frac{1}{\beta}\sqrt{1+\beta r^2}\,(\vec{\sigma}, \vec{p}) + 
\frac{(\vec{\sigma}, \vec{x})}{-\alpha +\sqrt{1+\beta r^2}}\,\,  \Big((\vec{x}, \vec{p}) -i \Big)\,.
\label{lastpsforstrangepot}
\end{eqnarray}

\section{An example of an exact solution of the 
Pauli-Schr\"{o}dinger equation}\label{example}
Usually the spin-orbital interaction term is treated perturbatively 
\cite{Merzbacher}. In the case of superintegrable systems we can 
obtain exact solutions. In this article we consider only one 
example: the potential $V_1(r) = \frac{1}{2r^2}$, $V_0 = V_0(r)$ 
and the integral of motion $X_P^1$ (\ref{solintofmotforpsone}). The 
system of equations to solve is 
\begin{eqnarray}
&& H \Psi = E \Psi\,, \label{schpaulieqforsol} \\
&& J^2 \Psi = j(j+1) \Psi\,, \label{sqroftotangforsol} \\
&& J_3 \Psi = m \Psi\,, \label{zcomofjforsol} \\
&& X_P^1 \Psi = \epsilon \Psi\,, \label{newintofmotionforsol}
\end{eqnarray}
where $\Psi = \Psi_{njm\epsilon} (r,\theta, \phi)$ is a two-component 
spinor. 

Equation (\ref{zcomofjforsol}) implies: 
\begin{eqnarray}
\Psi = \left(\begin{array}{c}
f_1(r, \theta)\, e^{i(m-\frac{1}{2})\phi} \\
f_2(r, \theta)\, e^{i(m+\frac{1}{2})\phi} \\
\end{array}\right)\,.
\label{spinorforsol}
\end{eqnarray}
Equation (\ref{newintofmotionforsol}) relates $f_1$ and $f_2$: 
\begin{eqnarray}
f_2(r, \theta) = \frac{\epsilon - \cos\theta}{\sin\theta} f_1(r, \theta)\,, \qquad \epsilon^2 = 1\,.
\label{f2relf1forsol}
\end{eqnarray}
Equation (\ref{sqroftotangforsol}) provides an equation for $f_1$, namely 
\begin{eqnarray}
f_{1, \theta\theta} + \frac{\epsilon}{\sin\theta}f_{1,\theta} - \left\{\frac{1}{\sin^2\theta} (m-\frac{1}{2})(m+\frac{1}{2} - \epsilon\cos\theta) -j(j+1) - \frac{1}{4} \right\} f_1 = 0\,, \label{deforangpartforsol} 
\end{eqnarray}
with
\begin{eqnarray}
j=\frac{1}{2},\frac{3}{2},\frac{5}{2},\ldots\,, \qquad m=\pm \frac{1}{2}, \pm \frac{3}{2}, \ldots, \pm j\,. \label{valforjandmforsol}
\end{eqnarray}

To solve (\ref{deforangpartforsol}) let us put 
\begin{eqnarray}
f_1 = R(r) F(\theta)\,, \label{sepforradandangpartforsol}
\end{eqnarray}
and obtain an equation for the angular part $F(\theta)$ from (\ref{deforangpartforsol}), 
namely 
\begin{eqnarray}
F_{\theta\theta} + \frac{\epsilon}{\sin\theta}F_{\theta} - \frac{1}{\sin^2\theta}\left\{m^2 - \frac{1}{4} -\epsilon (m - \frac{1}{2})\cos\theta -\Big(j(j+1) + \frac{1}{4}\Big) \sin^2\theta \right\} F = 0\,. \label{lastdeforangpartforsol} 
\end{eqnarray}
Equation (\ref{lastdeforangpartforsol}) can be solved in terms of 
Jacobi polynomials. We get different expressions for $m > 0$ and $m < 0$. 

For $m < 0$ we have
\begin{eqnarray}
F(\theta) = (1-z^2)^{\frac{1}{4} - \frac{m}{2}}\,P_{j+m}^{(\alpha,\beta)} (z)\,, \label{firstangpartsolaneg}
\end{eqnarray}
\begin{eqnarray}
\alpha = -m + \frac{\epsilon}{2}\,, \qquad \beta = -m - \frac{\epsilon}{2}\,, \qquad z = \cos\theta\,,
\end{eqnarray}
which is regular for $-1\leq z \leq 1$ (for $m < 0$).

For $m > 0$ we have
\begin{eqnarray}
F(\theta) = (1-z)^{\frac{m}{2} - \frac{\epsilon}{2} + \frac{1}{4}} (1+z)^{\frac{m}{2} + \frac{\epsilon}{2} + \frac{1}{4}}\,P_{j-m}^{(\alpha,\beta)} (z)\,, \label{secondangpartsolaneg}
\end{eqnarray}
\begin{eqnarray}
\alpha = m - \frac{\epsilon}{2}\,, \qquad \beta = m + \frac{\epsilon}{2}\,, \qquad z = \cos\theta\,.
\end{eqnarray}
Solution (\ref{secondangpartsolaneg}) is regular for $-1\leq z \leq 1$ (for $m > 0$).

Finally, to obtain the radial part of the solution we put the results 
obtained so far into (\ref{schpaulieqforsol}) and obtain the radial 
equation:
\begin{eqnarray}
-\frac{1}{2}\left(R^{\prime\prime} + \frac{2}{r} R^{\prime}\right) + \left\{V_0(r) + \Big(j(j+1) - \frac{3}{4}\Big)\frac{1}{2r^2}\right\}R = ER\,.
\label{radeqforsol}
\end{eqnarray}

We shall solve (\ref{radeqforsol}) for the case when 
$V_0(r)$ is the Coulomb potential 
\begin{eqnarray}
V_0(r) = \frac{\mu}{r}\,, \qquad \mu < 0\,.
\label{Coulombpotforsol}
\end{eqnarray}
The result is obtained in terms of Laguerre polynomials. 
We put 
\begin{eqnarray}
R(r) = e^{wr}r^p L(\sigma r)\,, 
\label{ansatz}
\end{eqnarray}
and obtain 
\begin{eqnarray}
\sigma r L^{\prime \prime} + 2(p+1+\frac{w}{\sigma} \sigma r) L^{\prime} + 
\left\{\frac{1}{\sigma r} \left(p(p+1) -j(j+1) + \frac{3}{4}\right) + \frac{2}{\sigma}(w(p+1) - \mu ) + \frac{w^2 + 2E}{\sigma^2} \sigma r \right\} L = 0\,.
\label{Lageqforsol}
\end{eqnarray}
This coincides with the equation for Laguerre polynomials 
$L = L_n^{\alpha} (\sigma r)$ if we put
\begin{eqnarray}
&& p = -\frac{1}{2} + \sqrt{j^2+j-\frac{1}{2}}\,, \qquad w = -\sqrt{-2E}\,, \qquad \sigma = 2\sqrt{-2E}\,, \nonumber \\
&& \frac{2}{\sigma} \Big(w(p+1) - \mu \Big) = n\,, \qquad \alpha = 2\, \sqrt{j^2+j-\frac{1}{2}}\,.
\label{parametersforsolforrad}
\end{eqnarray}

From (\ref{Lageqforsol}) we also find the bound state energies to be 
\begin{eqnarray}
E_{nj} = -\frac{\mu^2}{2\left(n+\frac{1}{2}+\sqrt{j^2+j-\frac{1}{2}}\right)^2}\,. 
\label{energyeigenvalues}
\end{eqnarray}

We see that the energy depends on only two quantum numbers, $n$ and 
$j$ whereas the wave function (\ref{spinorforsol}) depends on 
$n$, $j$, $m$ and $\epsilon$. The spin-orbital interaction 
removes the ``dynamical'' or ``accidental'' degeneracy with respect to the 
quantum number $j$. Superintegrability relates the two components 
of the spinor $\Psi_{njm\epsilon}$ and this made it possible to 
calculate the wave functions explicitly and exactly.


\section{Conclusions and outlook}
\label{conclusion}
The main results of this article can be summed up 
as a theorem.

\begin{proof}[{\bf Theorem 1.}]
The only first-order spherically symmetric 
superintegrable systems of type (\ref{generalhamiltonianwithhsec1}) in 
the Euclidean space $E_3$ are the following ones:
\begin{eqnarray}
\!\!\!\!\!\!\!\!\!\!\!\!\!\!\!\!\!\!\!\!\!\!\!\!\!\!\!\!\!\!\!\!\!\!\!\!\!\!\!\!\!\!\!\!\!\!\!\!\!\!\!\!\!\!\!\!\!\!\!\!\!\!\!\!\!\!\!\!
\!\!\!\!\!\!\!\!\!\!\!\!\!\!\!\!\!\!\!\!\!\!\!\!\!\!\!\!\!\!\!\!\!\!\!\!\!\!\!\!\!\!\!\!\!\!\!\!\!\!\!\!\!\!\!\!\!\!\!\!\!\!\!\!\!\!\!\!
\!\!\!\!\!\!\!\!\!\!\!\!\!\!\!\!\!\!\!\!\!\!\!\!\!\!\!\!\!\!\!\!\!\!\!\!\!\!\!\!\!\!\!\!\!\!\!\!\!\!\!\!\!\!\!\!\!\!\!\!\!
1.\qquad V_0 = \frac{\hbar^2}{r^2}\,, \qquad V_1 = \frac{\hbar}{r^2}\,.
\label{concleq1}
\end{eqnarray}
The integrals of motion are given in (\ref{generatorsforninedimalg}) 
and form the algebra (\ref{commutationrelationsforoneoverrsqr}), 
(\ref{algdirectsum}). The potentials (\ref{concleq1}) are induced 
from free motion by a gauge transformation.
\begin{eqnarray}
\!\!\!\!\!\!\!\!\!\!\!\!\!\!\!\!\!\!\!\!\!\!\!\!\!\!\!\!\!\!\!\!\!\!\!\!\!\!\!\!\!\!\!\!\!\!\!\!\!\!\!\!\!\!\!\!\!\!\!\!\!\!\!\!\!\!\!\!
\!\!\!\!\!\!\!\!\!\!\!\!\!\!\!\!\!\!\!\!\!\!\!\!\!\!\!\!\!\!\!\!\!\!\!\!\!\!\!\!\!\!\!\!\!\!\!\!\!\!\!\!\!\!\!\!\!\!\!\!\!\!\!\!\!\!\!\!
\!\!\!\!\!\!\!\!\!\!\!\!\!\!\!\!\!\!\!\!\!\!\!\!\!\!\!\!\!\!\!\!\!\!\!\!\!\!\!\!\!\!\!\!\!\!\!\!\!\!\!\!\!\!\!\!
2.\qquad V_0 = V_0(r)\,, \qquad V_1 = \frac{\hbar}{r^2}\,,
\label{concleq2}
\end{eqnarray}
where $V_0(r)$ is arbitrary. The integrals are given in 
(\ref{generatorsforsixdimalg}) and form the algebra 
(\ref{commutationrelationso3}), (\ref{algdirectsumo3o3}). The 
spin-orbital term $V_1$ is induced by a gauge transformation. 
\begin{eqnarray}
\!\!\!\!\!\!\!\!\!\!\!\!\!\!\!\!\!\!\!\!\!\!\!\!\!\!\!\!\!\!\!\!\!\!\!\!\!\!\!\!\!\!\!\!\!\!\!\!\!\!\!\!\!\!\!\!\!\!\!\!\!\!\!\!\!\!\!\!
\!\!\!\!\!\!\!\!\!\!\!\!\!\!\!\!\!\!\!\!\!\!\!\!\!\!\!\!\!\!\!\!\!\!\!\!\!\!\!\!\!\!\!\!\!\!\!\!\!\!\!\!\!\!\!\!\!\!\!\!\!\!\!\!\!\!\!\!
\!\!\!\!\!\!\!\!\!\!\!\!\!\!\!\!\!\!\!\!\!\!\!\!\!\!\!\!\!\!\!\!\!\!\!\!\!\!\!\!\!\!\!\!\!\!\!\!\!\!\!\!\!
3.\qquad V_0 = V_0(r)\,, \qquad V_1 = \frac{\hbar}{2r^2}\,,
\label{concleq3}
\end{eqnarray}
where $V_0(r)$ is arbitrary. The integrals are the two pseudoscalars 
(\ref{solintofmotforpsone}), (\ref{solintofmotforpstwo}) and the vector (\ref{solintofmotforvec}).
\begin{eqnarray}
\!\!\!\!\!\!\!\!\!\!\!\!\!\!\!\!\!\!\!\!\!\!\!\!\!\!\!\!\!\!\!\!\!\!\!\!\!\!\!\!\!\!\!\!\!\!\!\!\!\!\!\!\!\!\!\!\!\!\!\!\!\!\!\!\!\!\!\!
\!\!\!\!\!\!\!\!\!\!\!\!\!\!\!\!\!\!\!\!\!\!\!\!\!\!\!\!\!\!\!\!\!\!\!\!\!\!\!\!\!\!\!\!
4.\qquad V_0 = \hbar V_1\,, \qquad V_1=\frac{\hbar}{2r^2} \left(1+\frac{\alpha}{\sqrt{1+\beta r^2}}\right)\,, \qquad \alpha^2 = 1\,.
\label{concleq4}
\end{eqnarray}
The integral is the pseudoscalar (\ref{lastpsforstrangepot}). 

In all cases the components of the total angular momentum 
$J_i = L_i + \frac{\hbar}{2} \sigma_i$, $i=1,2,3$ 
are also integrals of motion.
\end{proof}

In the case of scalar particles ($V_1(r) = 0$) first-order 
superintegrability does not exist. The best known cases of 
second-order superintegrability are the Coulomb atom and 
the harmonic oscillator. In the first case the additional 
(to angular momentum) integrals of motion form a vector 
(the Laplace-Runge-Lenz vector). In the second case they form 
a two-valent tensor. In both cases the integrals generate a 
non Abelian Lie algebra and this leads to an additional degeneracy 
of the energy levels. 

In the case considered in this article the situation is different. 
First of all, first-order superintegrability does exist (see 
Theorem 1). For cases $3$ and $4$ of the Theorem, the additional pseudoscalar integrals commute with the total 
angular momentum. Hence it is possible to simultaneously 
diagonalize $H$, $J^2$, $J_3$ and the additional pseudoscalar integral $X$. 
In the example, considered in Section \ref{example}, the energy 
depends on two quantum numbers $n$ and $j$, the wave functions on 
$n$, $j$, $m$ and $\epsilon = \pm 1$. We thus have the geometric 
degeneracy related to the operator $J_3$ and also a discrete 
degeneracy due to $X$. 

Theorem 1 also provides examples of 
``pure quantum integrability'' \cite{Gravel.a, Gravel.b, Hietarinta}. 
The potentials $V_1$ and sometimes also the potentials $V_0$ 
disappear in the classical limit $\hbar \rightarrow 0$. 

It has been conjectured \cite{Tempesta} that all maximally superintegrable 
(scalar) systems are also exactly solvable. This means that their 
bound state energies can be calculated algebraically. Moreover their 
wave functions can be expressed as polynomials in the appropriate 
variables, multiplied by some overall factor. The example 
(\ref{schpaulieqforsol})-(\ref{newintofmotionforsol}) is 
superintegrable, but not maximally so. 
We have however shown that for $V_0 = \frac{\mu}{r}$ the system 
is exactly solvable. 

The conjecture of Ref. \cite{Tempesta} has been supported by many examples 
\cite{Tempesta, Rodriguez, Tremblay}.

In a future article we plan to study the potentials (\ref{concleq3}) 
and (\ref{concleq4}) in more detail, making other choices for 
$V_0$ in (\ref{concleq3}) and diagonalizing a more general operator 
$X_2 + \alpha X_1$ (see (\ref{solintofmotforpsone}) and 
(\ref{solintofmotforpstwo})). 

Another project that is being pursued is that of second-order 
superintegrability. The Hamiltonian is the same as in 
(\ref{generalhamiltonianwithhsec1}), however, the integrals, 
additional to total angular momentum  are not of the form 
(\ref{appendixmostgeneralintofmotion}) but are second-order polynomials 
in the linear momentum $\vec{p}$.


\section*{ACKNOWLEDGMENTS}
The authors thank A. S. Turbiner for a fruitful discussion. The research of P. W. was partly supported by a grant from NSERC of Canada. \.{I}. Y. acknowledges a postdoctoral fellowship awarded by the Laboratory of Mathematical Physics of the CRM, Universit\'{e} de Montr\'{e}al.

\renewcommand{\theequation}{A-\arabic{equation}}
\setcounter{equation}{0}  
\section*{APPENDIX}  
In this Appendix we give the full symmetric form 
of the integral of motions separately for scalars, 
pseudoscalars, vectors, axial vectors, tensors and 
pseudotensors. 

\noindent
{\bf{i) Scalars}}

The full symmetric form of ${X}_S$ is given as 
\begin{eqnarray}
X_S = f_1 + f_2\,(\vec{x}, \vec{p}) + f_3\,(\vec{\sigma}, \vec{L}) - \frac{i}{2}\,r\,f_2^{\prime} - \frac{3i}{2}\,f_2\,.
\label{appendixscalars}
\end{eqnarray}

\noindent
{\bf{ii) Pseudoscalars}}

The full symmetric form of ${X}_P$ could be given as 
\begin{eqnarray}
X_P = f_1\,(\vec{\sigma}, \vec{x}) + f_2\,(\vec{\sigma}, \vec{p}) + f_3\,(\vec{\sigma}, \vec{x})\,(\vec{x}, \vec{p}) - i\,\frac{(\vec{\sigma}, \vec{x})}{2r} 
\Big(f_2^{\prime} + r^2\, f_3^{\prime} + 4r\,f_3
\Big)\,.
\label{appendixpseudoscalars}
\end{eqnarray}

\noindent
{\bf{iii) Vectors}}

The full symmetric form of $\vec{X}_V$ can be written as 
\begin{eqnarray}
\vec{X}_V &=& \vec{x}\, \Big(f_1 - \frac{i}{2} \big(\frac{f_2^{\prime}}{r} + r f_3^{\prime} + 4f_3\big) + f_3\, (\vec{x}, \vec{p}) + f_4\, (\vec{\sigma}, \vec{L}) 
\Big) \nonumber \\
&+& f_2\, \vec{p} + f_6\, (\vec{\sigma} \wedge \vec{p}) + f_8\, (\vec{\sigma}, \vec{x})\,  \vec{L} \nonumber \\
&-& \frac{i}{2} (\vec{\sigma} \wedge \vec{x}) \Big(\frac{f_6^{\prime}}{r} + f_4- f_8 + rf_5^{\prime} + 4f_5 + 2i (f_5\, (\vec{x}, \vec{p}) + f_7)
\Big)\,.
\label{appendixvectors}
\end{eqnarray}

\noindent
{\bf{iv) Axial vectors}}

Let us now give the full symmetric form of 
$\vec{X}_A$
\begin{eqnarray}
\vec{X}_A &=& f_1\, \vec{L} + \vec{\sigma}\, \Big(f_2 - \frac{i}{2} \big( 3f_3 + r f_3^{\prime} + f_4 +f_6 \big) + f_3\, (\vec{x}, \vec{p})
\Big) \nonumber \\
&+& \vec{x}\, (\vec{\sigma}, \vec{x})\, \Big(f_5 - \frac{i}{2r} \big(f_4^{\prime} + f_6^{\prime}\big) - \frac{i}{2} \big(5f_7 + r f_7^{\prime}\big) + f_7\, (\vec{x}, \vec{p})
\Big) \nonumber \\ 
&+& f_4\, \vec{x}\, (\vec{\sigma}, \vec{p}) + f_6\, (\vec{\sigma}, \vec{x})\, \vec{p} \,.
\label{appendixaxialvectors}
\end{eqnarray}

\noindent
{\bf{v) Tensors}}

In the commutator relation $[H, X_T^{ik}]=0$, it is 
enough to consider only one component since the others 
then necessarily commute due to the rotations. Let us now 
give the full symmetric form of $X_T^{12}$
\begin{eqnarray}
X_T^{12} &=& xy \Big( f_1 + f_2 (\vec{x}, \vec{p}) - \frac{i}{2} (rf_2^{\prime} + 5f_2 ) + f_3 (\vec{\sigma}, \vec{L}) - i\frac{f_4^{\prime}}{r} 
\Big) \nonumber \\
&+& \big(zx\sigma_1 -zy\sigma_2 - (x^2 - y^2)\sigma_3\big)\Big(
\frac{i}{2} \big(f_3 - f_5 + rf_7^{\prime} + 5f_7 + \frac{1}{r}(f_8^{\prime} + f_9^{\prime})\big) - f_6 - f_7 (\vec{x}, \vec{p}) \Big) \nonumber \\
&+& f_4 (xp_y + yp_x) + f_5 (\vec{\sigma}, \vec{x}) (yL_1 +xL_2) + f_8 (y\sigma_2p_z - y\sigma_3p_y + x\sigma_3p_x -  x\sigma_1p_z) \nonumber \\
&-& f_9 \big(
(y\sigma_3 - z\sigma_2) p_y + (z\sigma_1 - x\sigma_3) p_x
\big) + f_{10} (\sigma_2 L_1 + \sigma_1 L_2 )\,.
\label{appendixtensor}
\end{eqnarray}

\noindent
{\bf{vi) Pseudotensors}}

Similarly it is enough to consider only one component. 
The full symmetric form of $X_Y^{12}$ is 
\begin{eqnarray}
X_Y^{12} &=& f_1 xy (\vec{\sigma}, \vec{p}) - xy (\vec{\sigma}, \vec{x}) \Big(
i\frac{f_1^{\prime}}{2r} - f_2 - f_3 (\vec{x}, \vec{p}) + i\frac{rf_3^{\prime}}{2} + 3if_3 + i\frac{f_4^{\prime}}{r}
\Big) \nonumber \\
&-& (y\sigma_1 +x\sigma_2)\Big(
\frac{i}{2} (f_1 + f_4 + rf_7^{\prime} - f_9) - f_6 - f_7 (\vec{x}, \vec{p}) + 2if_7 + i\frac{f_8^{\prime}}{2r}
\Big) \nonumber \\
&+& f_4 (\vec{\sigma}, \vec{x}) (xp_y + yp_x) + f_5 (yL_1 +xL_2) + f_8 (\sigma_1p_y + \sigma_2p_x) \nonumber \\
&+& f_9 \Big(
(x\sigma_3 - z\sigma_1) L_1 + (z\sigma_2 - y\sigma_3) L_2
\Big)\,.
\label{appendixpseudotensor}
\end{eqnarray}



\end{document}